\documentclass[12pt]{article}
\usepackage{graphicx}
\pdfoutput=1

\newcommand\pubnumber{CIPANP2018-Wietfeldt}
\newcommand\pubdate{\today}

\def\tulane{Department of Physics and Engineering Physics\\
Tulane University, New Orleans, LA 70118 USA}
\def\nist{National Institute of Standards and Technology\\
Gaithersburg, MD 20899 USA}
\def\hamilton{Physics Department\\
Hamilton College, Clinton, NY 13323 USA}
\def\indiana{CEEM\\
Indiana University, Bloomington, IN 47408 USA}
\def\depauw{Department of Physics and Astronomy\\
DePauw University, Greencastle, IN 46135 USA}

\def\Title#1{\begin{center} {\Large #1 } \end{center}}
\def\Author#1{\begin{center}{ \sc #1} \end{center}}
\def\Address#1{\begin{center}{ \it #1} \end{center}}

\newcommand\pubblock{\rightline{\begin{tabular}{l} \pubnumber\\
         \pubdate  \end{tabular}}}
\newenvironment{Abstract}{\begin{quotation}  }{\end{quotation}}
\newenvironment{Presented}{\begin{quotation} \begin{center} 
             PRESENTED AT\end{center}\bigskip 
      \begin{center}\begin{large}}{\end{large}\end{center} \end{quotation}}
\def\Acknowledgements{\bigskip  \bigskip \begin{center} \begin{large}
             \bf ACKNOWLEDGEMENTS \end{large}\end{center}}

\def\beq{\begin{equation}}
\def\eeq#1{\label{#1}\end{equation}}
\def\eeqn{\end{equation}}

\def\beqa{\begin{eqnarray}}
\def\eeqa#1{\label{#1}\end{eqnarray}}
\def\eeqan{\end{eqnarray}}

\let\bar=\overbar

\def\Dslash{\not{\hbox{\kern-4pt $D$}}}
\def\dslash{\not{\hbox{\kern-2pt $\del$}}}

\def\msb{{\bar{\ssstyle M \kern -1pt S}}}

\begin{document}
\begin{titlepage}
\pubblock

\vfill
\Title{Measurement of the electron-antineutrino correlation in neutron beta decay: aCORN experiment}
\vfill
\Author{ F.~E.~Wietfeldt, W.~A.~Byron\footnote{current address: University of Washington, Seattle, WA USA}, 
G.~Darius, C.~R.~DeAngelis,  M.~T.~Hassan\footnote{current address: NIST, Gaithersburg, MD USA}}
\Address{\tulane}
\Author{ M.~S.~Dewey, M.~P.~Mendenhall\footnote{current address: Lawrence Livermore National Laboratory, Livermore, CA USA}, 
J.~S.~Nico}
\Address{\nist}
\Author{ B.~Collett, G.~L.~Jones}
\Address{\hamilton}
\Author{ A.~Komives}
\Address{\depauw}
\Author{ E.~J.~Stephenson}
\Address{\indiana}
\vfill
\begin{Abstract}
The aCORN experiment uses a novel asymmetry method to measure the electron-antineutrino correlation ($a$-coefficient) in free neutron decay that does not require precision proton spectroscopy. aCORN completed two physics runs at the NIST Center for Neutron Research. The first run on the NG-6 beam line in 2013--2014 obtained the result $a$ = 0.1090 +/- 0.0030 (stat) +/- 0.0028 (sys), a total uncertainty of 3.8\%. The second run on the new NG-C high flux beam line promises an improvement in precision to $<2$\%.
\end{Abstract}
\vfill
\begin{Presented}
CIPANP2018
\end{Presented}
\vfill
\end{titlepage}
\def\thefootnote{\fnsymbol{footnote}}
\setcounter{footnote}{0}

\section{Introduction}

The main experimental observables of neutron beta decay are described by the formula of Jackson, Treiman, and Wyld \cite{JTW57}, derived from the Hamiltonian for $J=\frac{1}{2}$ to $\frac{1}{2}$ beta decay
\begin{equation}
\label{E:JTWeqn}
N \propto \frac{1}{\tau_n}F(E_e) \left[ 1 
+ a\frac{\mbox{\boldmath $p_e$}\cdot \mbox{\boldmath $p_\nu$}}{E_e E_\nu} + b\frac{m_e}{E_e}
+ \mbox{\boldmath $\mathcal P$}\cdot \left( A\frac{\mbox{\boldmath $p_e$}}{E_e} + B\frac{\mbox{\boldmath $p_\nu$}}{E_\nu}
+ D\frac{(\mbox{\boldmath $p_e$}\times \mbox{\boldmath $p_\nu$})}{E_e E_\nu} \right) \right].
\end{equation}
Here $E_e$, \mbox{\boldmath $p_e$}, $m_e$ are the beta electron total energy, momentum, and mass; \mbox{\boldmath $p_{\nu}$}, $E_{\nu}$ are
is the antineutrino momentum and energy; and $F( E_e)$ is the beta energy spectrum. The neutron decay lifetime is $\tau_n$ and the parameters $a$, $A$, $B$, and $D$ are correlation coefficients that are experimentally measured. It is assumed here that the neutrons are in a spin 
polarization state \mbox{\boldmath $\mathcal P$}, while the beta electron and antineutrino spins are averaged over, which is typically the case
in an experiment. In the Standard ($V-A$) Electroweak Model, neglecting recoil order corrections, the values of these correlation coefficients and the lifetime 
are related to two basic parameters in the theory: 
the weak vector and axial vector coupling constants $G_V$ and $G_A$. If we write their ratio as $G_A / G_V = \lambda$, we have \cite{JTW57}
\[
\tau_n =  \left(\frac{2 \pi^3 \hbar^7}{m_e^5 c^4 f}\right)\frac{1}{G_V^2 + 3 G_A^2} \qquad 
a  =  \frac{1 - \lambda^2}{1 + 3\lambda^2} 
\]
\begin{equation}
\label{E:SMtaABD}
A  =  -2\frac{{\rm Re}\{\lambda \} + \lambda^2}{1 + 3\lambda^2} \qquad
B  =  -2\frac{{\rm Re}\{\lambda \} - \lambda^2}{1 + 3\lambda^2} \qquad
D  =  2\frac{{\rm Im}\{\lambda \}}{1 + 3\lambda^2}.
\end{equation}
The quantity $f$ in $\tau_n$ is a calculable phase space factor. A measurement of $\tau_n$ plus any one of $a$, $A$, or $B$ gives $G_A$ and $G_V$
uniquely. Multiple precision measurements of these parameters overconstrain the system and can be used test the
self-consistency of the Standard Electroweak Model and search for hints of new physics.\par
Bloch and M{\o}ller proposed  \cite{BM35} that the electron-neutrino angular correlation, the $a$-coefficient in equations \ref{E:JTWeqn} and \ref{E:SMtaABD}, be used to experimentally distinguish between scalar, vector, axial vector, and tensor interactions in beta decay. Scalar and tensor interactions require the beta electron and antineutrino to be emitted in the same helicity state while vector and axial vector require opposite helicities. Therefore, in allowed Fermi decay where the leptons are produced in a spin singlet with $L=0$, the scalar will produce a negative (specifically $a = -1$) and the vector a positive ($a = +1$) angular correlation. Similarly, in Gamow-Teller decay the leptons are produced in a spin triplet with $L=0$, so the axial vector will give a negative ($a = -1/3$) and the tensor a positive ($a = +1/3$) angular correlation. For mixed decays, such as neutron decay, $a$ will take an intermediate value. This property of the $a$-coefficient was exploited in the 1950's to demonstrate the $V-A$ nature of 
the weak force \cite{Al59}. The data clearly favored an interaction that is predominantly $V,A$, although the possibility of much weaker $S,T$ interactions, which could be introduced by high mass scale physics beyond the Standard Model, could not be ruled out. 
The best current limits on $S,T$ weak interactions, from nuclear and pion beta decay, are at the $10^{-3}$ level \cite{Har15, Byc09, Sev06, Cir13, Vos15}. Improved measurement of the neutron $a$-coefficient will help push limits on $S,T$ weak interactions to higher sensitivity.
\section{The aCORN method}
\begin{figure}[t]
\begin{center}
\includegraphics[width = 4in]{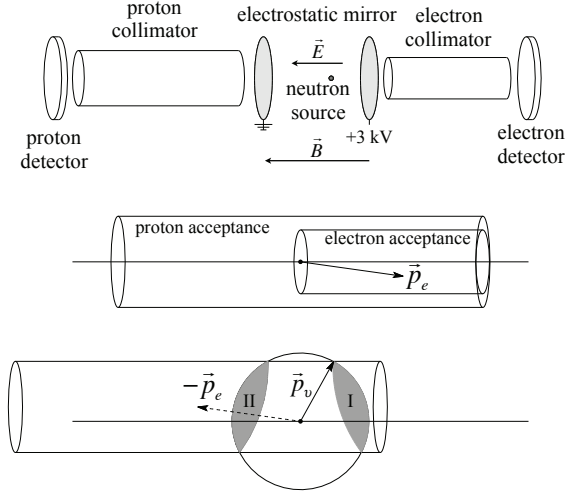}
\end{center}
\vspace{-0.25in}
\caption{\label{F:cylfig} The aCORN method, illustrated here for the case where the decay vertex is on the experimental axis. Beta electrons are
accepted up to a maximum transverse momentum set by the electron collimator radius and the axial magnetic field strength, with axial momentum
toward the electron detector (top), represented as a cylinder in momentum space (middle). The recoil proton momentum acceptance is also a cylinder, but due to the
electrostatic mirror all axial momenta are accepted. The bottom figure shows the momentum acceptance of the antineutrino, when the
electron and proton were detected in coincidence. By conservation of momentum and energy this is the intersection of a cylinder and the surface of a sphere, defining
two regions marked I and II. Region I is correlated with the electron momentum and region II is anticorrelated.}
\end{figure}

Previous neutron $a$-coefficient experiments relied on measuring the shape of the recoil proton energy spectrum and were systematically limited at about 5\% relative uncertainty \cite{Gri68, Str78, Byr02}. This approach is statistically more powerful than a coincidence experiment but requires precision low energy (751 eV maximum) proton spectroscopy, a difficult experimental challenge. aCORN is different; it employs a novel ``wishbone asymmetry'' method first proposed by Yerozolimsky and Mostovoy \cite{Bal94, Yer04, Wie05, Wie09}. The basic scheme is illustrated in figure \ref{F:cylfig} (top).\par
Consider a neutron decay vertex on the axis of a long solenoidal magnet. The beta electron and recoil proton are transported by the magnetic field to electron and proton detectors on the ends. Electron and proton collimators restrict the transverse momenta of decay particles that are detected. The decay region is surrounded by a 3 kV electrostatic mirror that reflects all recoil protons toward the proton detector.  It also preaccelerates all protons to similar velocities to reduce sensitivity to systematic effects associated with transverse magnetic fields and residual gas interactions. A momentum space representation of this scheme is shown in figure \ref{F:cylfig} (middle). With our assumption that the decay vertex was on axis, the acceptances in momentum space are cylinders as shown; for each particle there is a maximum transverse momentum defined by the collimator radius and axial magnetic field. The axial momenta of detected electrons must be toward the electron detector, but the electrostatic mirror allows all proton axial momenta to be accepted. The momentum vector for a particular electron that is detected is shown as $\vec{p_e}$. If the associated recoil proton momentum is within its acceptance cylinder, an electron-proton coincidence event is counted. The antineutrino is not detected, but because the cold neutrons decay effectively at rest,  the antineutrino momentum satisfies $\vec{p_{\nu}} \approx -\vec{p_e} - \vec{p_p}$ and the antineutrino momentum acceptance is the cylinder shown in figure \ref{F:cylfig} (bottom), constructed by subtracting the proton cylinder from $-\vec{p_e}$. The detected electron momentum fixes the electron energy, and the proton kinetic energy can be (to good approximation) neglected, therefore the antineutrino energy must satisfy $E_{\nu} \approx Q_{\beta} - E_e$ and its momentum lies on the surface of the sphere shown in the bottom figure. For neutron decay electrons and protons that are detected in coincidence, conservation of energy and momentum confines the antineutrino momentum to the intersection of the cone and the surface of this sphere, indicated by the shaded regions I and II. These regions serve as virtual neutrino detectors. By construction they subtend equal solid angle from the origin of momentum space. Antineutrino momenta associated with region I are correlated with the electron momentum, and those associated with region II are anticorrelated, so the asymmetry in event rates associated with the two regions measures the $a$-coefficient. When the decay vertex is off-axis, as in the case of a beam source, the picture is somewhat more complicated -- the momentum acceptance cylinders are elliptical rather than circular -- but the construction is similar and conclusions are the same.
\par
\begin{figure}[t]
\begin{center}
\includegraphics[width = 3.5in]{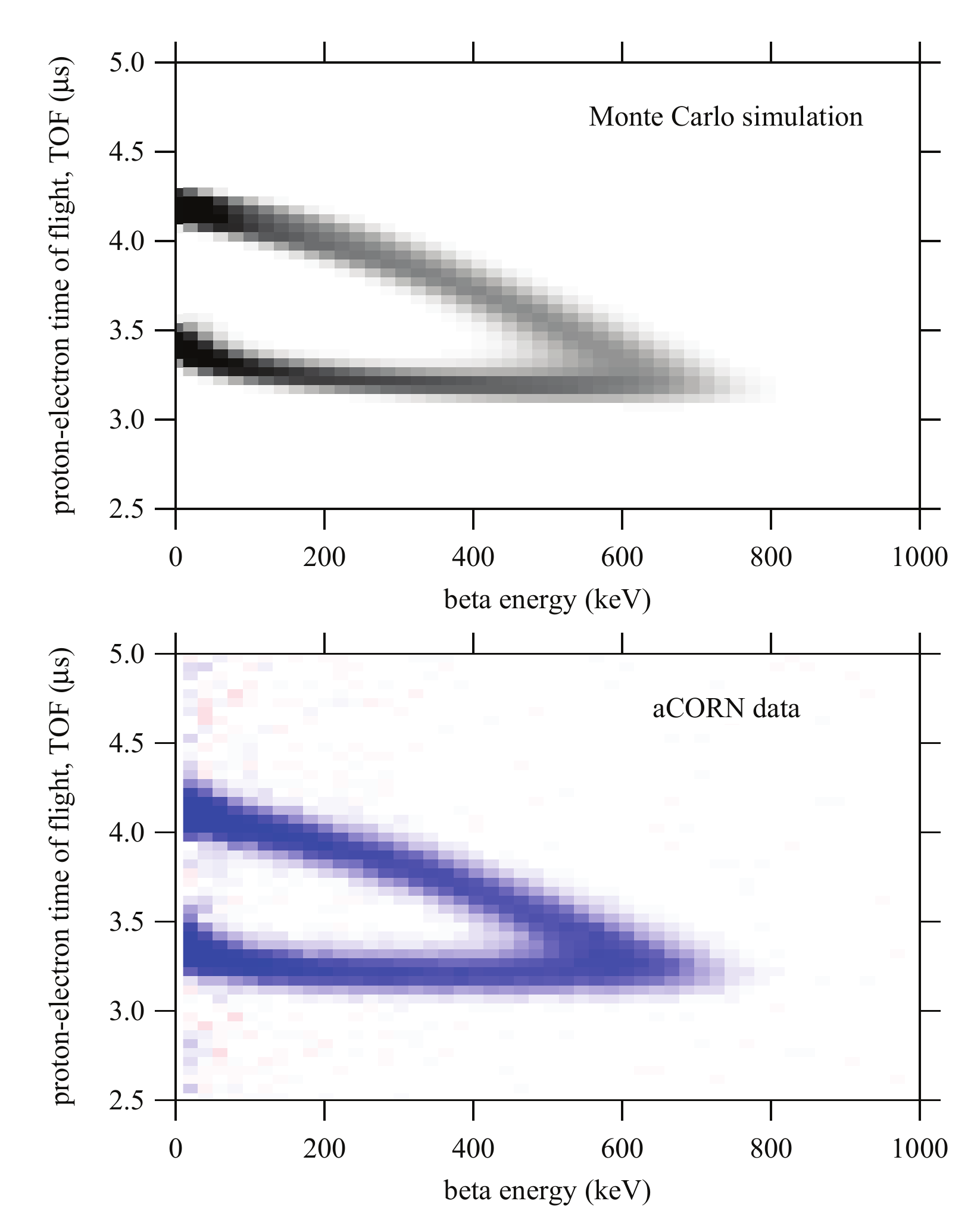}
\end{center}
\vspace{-0.25in}
\caption{\label{F:wishbones} The aCORN ``wishbone'' plot of proton time of flight vs. beta energy for neutron decay events. The top plot is a Monte Carlo simulation and the bottom is a sample (about 400 hours) of aCORN data. Blue pixels are positive and red
are negative (due to the background subtraction)}
\end{figure}

In the aCORN experiment we measure the electron energy and the proton time-of-flight (TOF), the time between electron and proton detection, for coincidence events. Neutron decays form a characteristic ``wishbone'' distribution shown in figure \ref{F:wishbones}. The lower branch  containing faster protons corresponds to the shaded region I in figure \ref{F:cylfig} and the upper branch containing slower protons corresponds to region II. The gap between the branches corresponds to the kinematically forbidden gap between regions I and II on the antineutrino sphere. We obtain, after many decays, $N_{\rm I}(E)$ events in group I (fast proton branch) and $N_{\rm II}(E)$ events in group II (slow proton branch) for a vertical slice of the wishbone with electron energy $E$.
Using equation \ref{E:JTWeqn}, with neutron polarization \mbox{\boldmath $\mathcal P$ = 0}, we have
\begin{equation}
N^{I(II)}(E) = F(E) \int\int \left( 1 + a v \cos\theta_{e\nu} \right) d\Omega_e\, d\Omega^{I(II)}_{\nu},
\end{equation}
where $F(E)$ is the beta energy spectrum, $v$ is the beta velocity (in units of $c$), $\cos\theta_{e\nu}$ is the cosine of the angle between the electron and antineutrino momenta, and $d\Omega_e$, $d\Omega^{I(II)}_{\nu}$ are elements of solid angle of the electron and antineutrino (group I, II) momenta. The integrals are taken over the momentum acceptances shown in figure \ref{F:cylfig}. Given that, by design, the total solid angle products are equal for the two groups: 
$\Omega_e\,\Omega^I_{\nu}$ = $\Omega_e\,\Omega^{II}_{\nu}$, it is straightforward to show that the $a$-coefficient  is related to the wishbone asymmetry $X(E)$:
\begin{equation}
\label{E:asym}
X(E) = \frac{N^I(E) - N^{II}(E)}{N^I(E) + N^{II}(E)} = \frac{ \frac{1}{2} a v \left( \phi^I(E) - \phi^{II}(E) \right)}{1 + \frac{1}{2} a v \left( \phi^I(E) + \phi^{II}(E) \right)}
\end{equation}
The parameters $\phi^I(E)$ and  $\phi^{II}(E)$ are defined by
\begin{equation}
\label{E:phis}
\phi^I(E) = \frac{ \int d\Omega_e \int_I d\Omega_{\nu} \cos\theta_{e\nu} }{\Omega_e \Omega_{\nu}^I} \qquad
\phi^{II}(E) = \frac{ \int d\Omega_e \int_{II} d\Omega_{\nu} \cos\theta_{e\nu} }{\Omega_e \Omega_{\nu}^{II}}.
\end{equation}
Note that $\phi^I(E)$ and $\phi^{II}(E)$ can be understood as the average value of $\cos\theta_{e\nu}$ for the coincidence detection acceptances associated with each wishbone branch.  They are simply geometrical factors; they contain no physics and in particular they do not depend on the value of the $a$-coefficient. They are functions of the transverse momentum acceptances of the proton and electron so they can be precisely calculated from the known axial magnetic field and collimator geometries. 
\par
The second term in the denominator of equation \ref{E:asym} has a numerical value less than 0.005 in the energy range of interest (100--360 keV), so we can treat it as a first order correction and write
\begin{equation}
\label{E:aEffective}
X(E) = a f_a(E)\left[ 1 + \delta_1(E) \right] + \delta_2(E)
\end{equation}
with
\begin{equation}
\label{E:faE}
f_a(E) = \frac{1}{2} a v \left( \phi^I(E) - \phi^{II}(E) \right)
\end{equation}
and 
\begin{equation}
\delta_1(E) = -\frac{1}{2} a v \left( \phi^I(E) + \phi^{II}(E) \right).
\end{equation}
There is another correction that comes from our neglect of the proton's kinetic energy in the momentum space discussion of figure \ref{F:cylfig}. If we account for this energy, the antineutrino sphere is slightly oblong and the solid angles of groups I and II differ by approximately 0.1\%. This produces a small ($<10^{-3}$) intrinsic asymmetry that is independent of the $a$-coefficient, represented by $\delta_2(E)$ in equation \ref{E:aEffective}, easily computed by Monte Carlo. Omitting the small corrections we see that $X(E) = a f_a(E)$; the experimental wishbone asymmetry is proportional to the $a$-coefficient and the geometric function $f_a(E)$.
\section{Experiment and Results}
\begin{figure}
\begin{center}
\includegraphics[width = 4.0in]{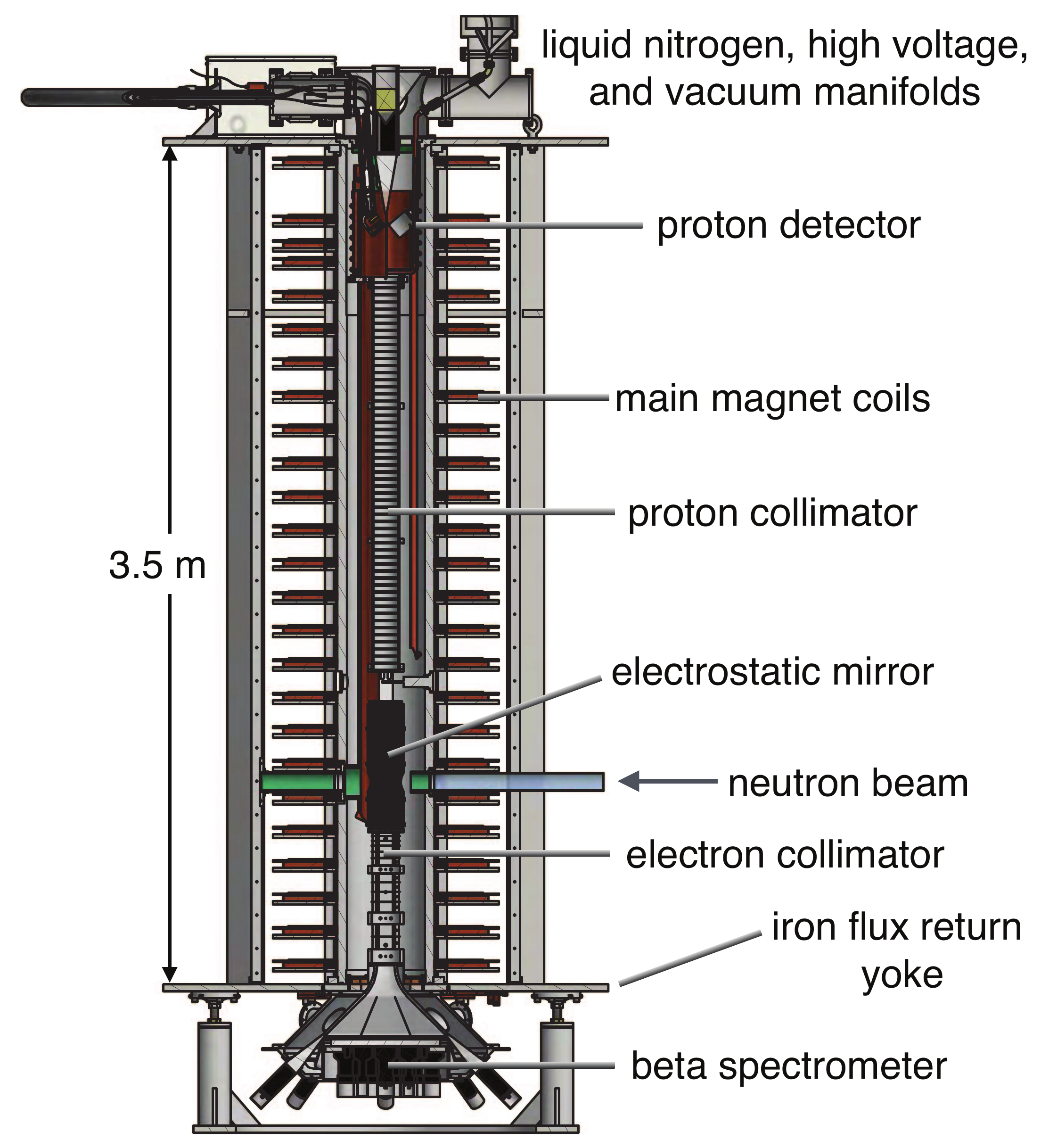}
\end{center}
\vspace{-0.25in}
\caption{\label{F:tower} A diagram of the aCORN apparatus showing the major components and arrangement.}
\end{figure}

The aCORN experiment was installed and operated on the end position NG-6 at the National Institute of Standards and Technology (NIST) Center for Neutron Research (NCNR) \cite{NCNR} from February 2013 to May 2014, collecting 1900 beam hours of physics data.  A diagram of the aCORN apparatus is shown in figure \ref{F:tower}. The main magnet was a set of 25 water-cooled flat coils that produced a 36.2 mT axial magnetic field. Sets of 25 axial trim coils and 45 transverse trim coils, each independently served by computer-controlled current supplies, were used to reduce transverse magnetic fields to less than 0.004 mT in the electrostatic mirror and proton collimator. 
\par
The electrostatic mirror consisted of a 0.25 mm wall PTFE cylindrical tube electroplated with 4.5 $\mu$m of copper on the inner surface. The copper was divided into 63 precise horizontal bands by photolithography and connected to a chain of 1.0 M$\Omega$ resistors to produce an approximately linearly varying electrostatic potential on the wall. At the top and bottom of the cylinder were wire grid planes (linear arrays of 100 $\mu$m wire, 2.0 mm spacing) held at ground and +3 kV, respectively. 
Below the electrostatic mirror was the beta collimator, a series of 17 tungsten plates, 0.5 mm thick with 5.5 cm diameter apertures, unevenly spaced to minimize the number of scattered electrons that enter the beta spectrometer. The proton collimator was a 140 cm long monolithic aluminum tube containing a series of 55 precision turned 8 cm diameter knife-edge apertures. It was sufficiently long that all neutron decay protons made at least one full cycle of helical motion within it before reaching the proton detector. The backscatter suppressed beta spectrometer was attached to the bottom of the tower, below the electron collimator. It consisted of a 5 mm thick, circular piece of Bicron BC-408 plastic scintillator \cite{DISCLM} viewed by a hexagonal array of 19 photomultiplier tubes. A set of eight plastic scintillator veto counters were employed to suppress events where the beta electron backscattered from the main detector without depositing its full energy. Further details on the design and operation of the beta spectrometer can be found in a previous publication \cite{Has17}. The proton detector was a 600 mm$^2$ liquid-nitrogen cooled silicon surface barrier detector, and a set of focusing electrodes, held at -29 kV and mounted slightly off axis so that electrons with upward trajectories cannot backscatter from it and subsequently reach the beta spectrometer. 
\par
Figure \ref{F:wishbones} (bottom) shows the background subtracted and deadtime corrected wishbone plot from a typical data set (about 400 beam hours). Neutron beam-induced background was significant; the coincidence signal to background ratio was typically about 0.4 in the wishbone region. Data were collected so that each electron signal that arrived within 10 $\mu$s before or 1 $\mu$s after each proton signal was treated as a separate event, which guaranteed that the spectrum of random coincidences associated with background was flat in the TOF domain, within statistical fluctuations, enabling a very clean background subtraction.  The wishbone asymmetry $X(E)$ was measured in the energy range 100--360 keV.  Below 100 keV electrons may miss the active region of the scintillator detector, complicating the calculation of the geometric acceptance function $f_a(E)$. Above 360 keV the wishbone branches overlap, obscuring the asymmetry. 
\par
\begin{figure}[h]
\begin{center}
\includegraphics[width = 4.0in]{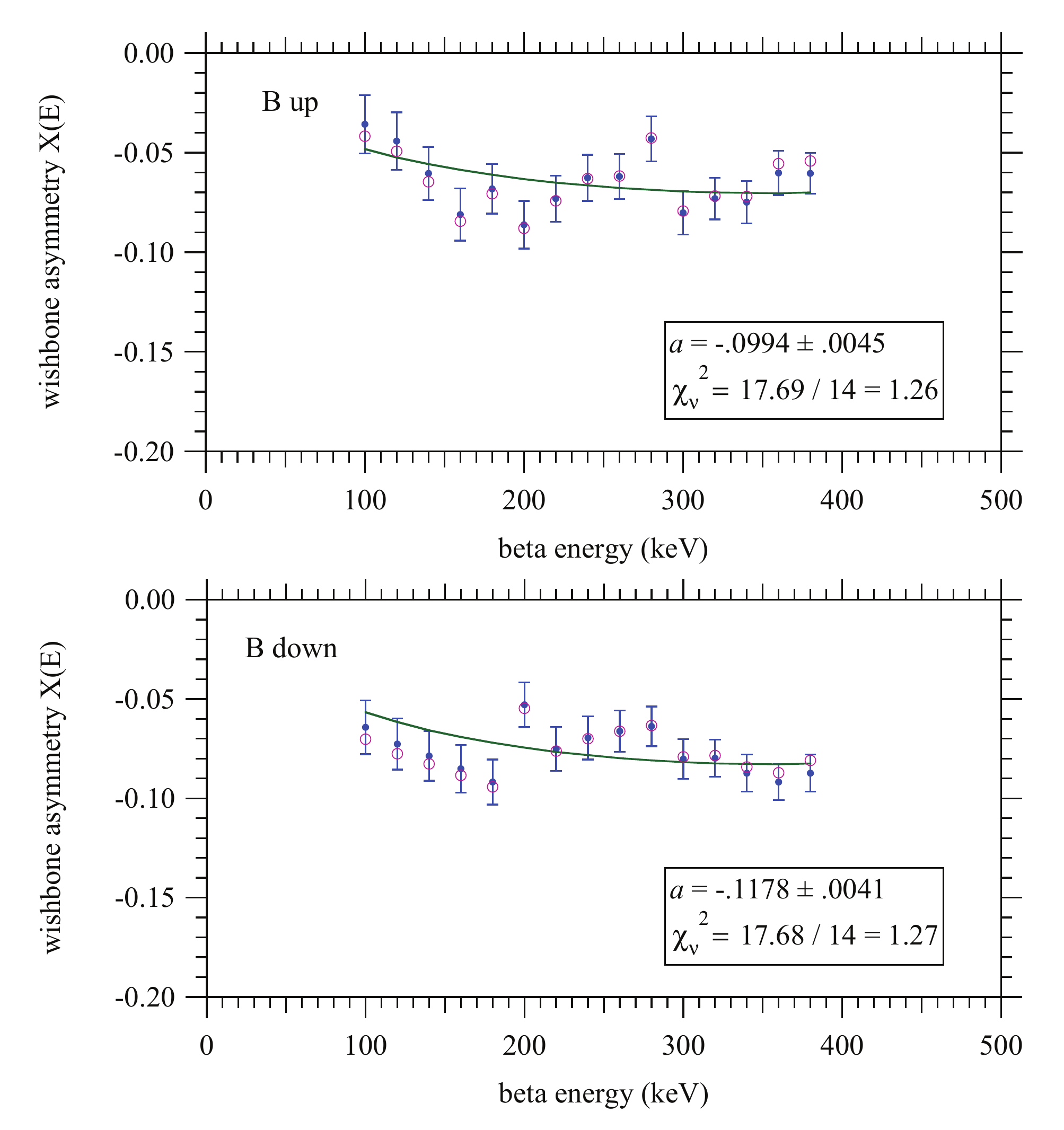}
\end{center}
\vspace{-0.25in}
\caption{\label{F:aFits} Open circles: The measured, uncorrected, wishbone asymmetry $X(E)$ for each magnetic field direction. Solid points: The same data including the corrections $\delta_1(E)$,  $\delta_2(E)$, and the energy-dependent systematic corrections. Error bars are statistical uncertainty. Solid curves: The product $a f_a(E)$, where $a$ is the best fit value of the $a$ coefficient in each case.}
\end{figure}

Figure \ref{F:aFits} shows the measured wishbone asymmetry $X(E)$ for the full data set for each magnetic field direction. Open circles are uncorrected data.  Solid circles include the calculated energy-dependent corrections for $\delta_1(E)$ and $\delta_2(E)$, and some energy-dependent systematic corrections.  Also shown is the function $f_a(E)$ multiplied by the best-fit value of the $a$ coefficient for each field direction. We attribute the difference in asymmetry to a slight residual neutron beam polarization of approximately $P \approx 0.006$. Taking the simple average eliminates this effect. Our result from the NG-6 run is \mbox{$a = -0.1090 \pm 0.0030\mbox{(stat)}\pm 0.0028\mbox{(sys)}$}. Additional details on the design, construction, alignment, and calibration of the aCORN apparatus and individual components, and analysis of systematic effects, can be found in previous publications \cite{Wie09,Col17,Has17,Dar17}.
\par
In 2014 the aCORN experiment was moved to the new high-flux end position NG-C at the NCNR, where it ran until 2016. The NG-C data set is about ten times larger and is expected to produce a result with $< 2\%$ relative uncertainty in the $a$-coefficient. Analysis of those data is in progress.
\newpage

\Acknowledgements
This work was supported by the National Science Foundation, U.S. Department of Energy Office of Science, and NIST (US Department of Commerce). We thank the NCNR for providing the neutron facilities used in this work, and for technical support, especially Eli Baltic, George Baltic, and the NCNR Research Facilities Operations Group.


\begin{thebibliography}{99}
\bibitem{DISCLM} Certain trade names and company products are 
mentioned in the text or identified in illustrations in order to 
adequately specify the experimental procedure and equipment used. In 
no case does such identification imply recommendation or endorsement 
by the National Institute of Standards and Technology, nor does it imply 
that the products are necessarily the best available for the purpose.
\end{thebibliography}
\end{document}